\documentstyle[preprint,aps]{revtex}

\input{psfig}
\begin{document}

\title{Event Anisotropy in High Energy Nucleus-Nucleus Collisions}

\author{H. Liu$^{(1)}$, S. Panitkin$^{(1)}$, and N. Xu$^{(2)}$}

\address{
$^{(1)}$Kent State University, Kent, Ohio 44242 \\
$^{(2)}$Lawrence Berkeley National Laboratory, Berkeley, California 94720\\
}
\vspace{0.2in} 

\date{June 29, 1998}
\maketitle

\begin{abstract}
The predictions of event anisotropy parameters from transport model RQMD
are compared with the recent experimental measurements for 158$A$ GeV
Pb+Pb collisions. Using the same model, we study the time evolution of
event anisotropy at 2$A$ GeV and 158$A$ GeV for several colliding
systems. For the first time, both momentum and configuration space
information are studied using the Fourier analysis of the azimuthal
angular distribution. We find that, in the model, the initial geometry
of the collision plays a dominant role in determining the anisotropy
parameters.
\end{abstract}

\narrowtext

Event anisotropy, often called flow, has been observed in heavy ion
collisions at every laboratory energy
\cite{Partlan95,Crochet98,Liu97,Barrette97,ags,Poskanzer97,NA49flow,WA98flow,Ritter97}.
It is believed that the information about the equation of state (EOS)
can be obtained from the study of flow~\cite{Csernai94}. Recently,
several authors argue that the event shape with respect to the
reaction plane may carry the information about pressure created at the
early stage of the collision
\cite{Sorge97,Sorge971,Rischke95,Gyulassy97,Danielewicz98}.

Theoretically, in terms of hydrodynamics, the connection between the
EOS and flow is well defined and has been studied extensively for many
years. This is especially the case for collisions at beam energy below
15$A$ GeV. (For the latest review of the subject, see
Ref.~\cite{Ritter97,Danielewicz95}.) However, it is still not clear
whether the experimentally observed event anisotropy is of a dynamic
origin (pressure-driven hydroexpansion) or due to shadowing of cold
nuclear matter, passing time etc. If the initial geometry of the
system plays an important role, the information about the collision
dynamics may be obscured.  In any case, it seems to be necessary to
understand the interplay between these two competing effects.  For
this purpose, we use the transport model RQMD(v2.3 cascade
mode)~\cite{Sorge95} to study the event anisotropy as a function of
the colliding system and beam energy. We also investigate the
collision time dependence of the event shapes.


In RQMD, the reaction plane is defined by the impact parameter ($x$
direction) and the projectile momentum ($z$ direction). Particle
azimuthal distribution with respect to the reaction plane at a given
rapidity window can be deconvoluted by the Fourier expansion
~\cite{Voloshin96},

\begin{equation}
\frac{dN}{d\phi} \approx v_0 ( 1 + 2 v_1\cos(\phi) + 2 v_2\cos(2\phi))
\end{equation}

\noindent
where the first and second Fourier coefficients, $v_1$ and $v_2$, are
connected to the directed flow and elliptic flow,
respectively~\cite{Voloshin96}. The coefficient $v_0$ is a
normalization constant and $\phi$ is defined as the azimuthal angle
difference between the particle and the reaction plane. In our case,
the reaction plane angle is explicitly set to zero. Hence $\phi$ angle
in momentum space is $\phi = {\rm tan^{-1}}(p_y/p_x)$, and in
configuration space $\phi = {\rm tan^{-1}}(y/x)$. At a given
rapidity, the coefficients are:

\begin{equation}
v_1 = \langle cos(\phi) \rangle, \hspace{0.25in}
v_2 = \langle cos(2\phi) \rangle.
\end{equation}

In the following, we will first compare the recent NA49 measurements
of Fourier moments with RQMD calculations. In section II and III, the
colliding system dependence and time evolution of anisotropy
parameters $v_1$ and $v_2$ will be discussed. Finally, a brief summary
will be presented in section IV.

\section{ Comparison of data with model predictions for 158$A$ GeV
Pb+Pb collisions}

\begin{center}
\begin{figure}[c]
\centerline{\psfig{file=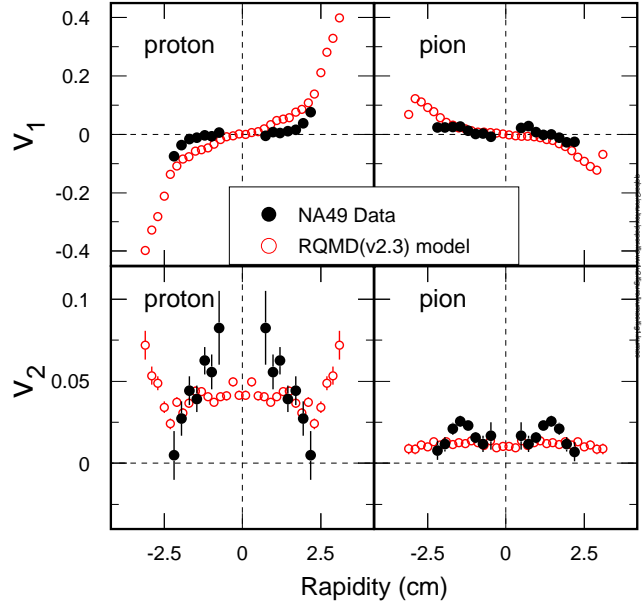,width=10.5cm}}
\caption{ $v_1, v_2$ for proton and pion as a function of rapidity (filled
circles - experimental results, open circles - model calculations) from
158$A$ GeV Pb+Pb collisions. }
\label{fig1}
\end{figure}
\end{center}

The experimentally measured ($v_1, v_2$) parameters for proton and
pion as a function of rapidity are shown as filled circles in
Figure~\ref{fig1}. These data are extracted from the 158$A$ GeV Pb+Pb
collisions by the NA49 collaboration~\cite{Poskanzer97,NA49flow}.
More details about methods involved in the experimental determination
of these parameters can be found in the recent paper ~\cite{pv98}.
The model calculations (open circles) are done within similar impact
parameter range and acceptance ($p_t,y$) cuts. As one can see, the
model calculations are in reasonable agreement with the data for both
protons and pions. The sign of the first Fourier moment $v_1$ for
pions is opposite to that of protons for both data and model
calculations indicating the nuclear shadowing effects in the
collision. An interesting feature in Figure~\ref{fig1} is that there
seems to be two slopes for the function $v_1(y)$: at $|y|\leq$
1.5, the distribution is relatively flatter compared to the region
$|y|>$ 1.5. This implies that the physics around the
mid-rapidity region is different from that near the projectile
(target) rapidity region. One possible interpretation is that near the
beam rapidity region, the number of rescatterings is small, therefore
the initial geometry leads to a large value of $v_1$. On the contrary,
a large number of rescatterings washes out the initial memory and
leaves small anisotropy around the mid-rapidity.

As predicted in Ref.~\cite{Ollitrault92}, the second Fourier moment
$v_2$ is found to be positive for both nucleons and pions at this
beam energy. This implies an enhancement of in-plane emission from the
particle source. More discussions on this point will be presented
later in this paper.

Since the model predictions are in reasonable agreement with the
experimental data, it may be instructive to study system size
dependence and the space-time evolution of the anisotropy parameters
within the model.

\section{ \small{$v_1$} for different collision systems}
 
\begin{center}
\begin{figure}[c]
\centerline{\psfig{file=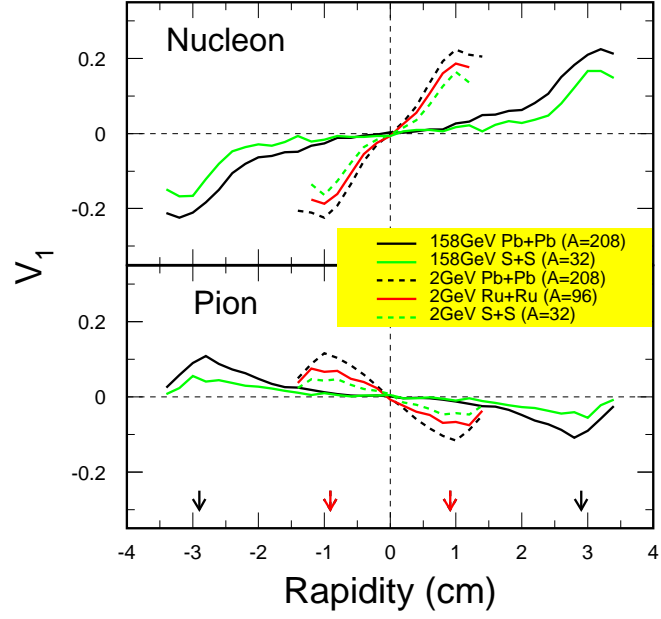,width=10.5cm}}
\caption{ $v_1$ as a function of rapidity $y$, in center-of-mass
frame, for nucleons and pions for different colliding systems. The arrows
indicate the beam and target rapidities.}
\label{fig2}
\end{figure}
\end{center}

In Figure~\ref{fig2}, the model predicted $v_1$ for nucleons and pions
as a function of rapidity are shown for 2$A$ GeV S+S ($3\leq b\leq5$
fm), 2$A$ GeV Ru+Ru \footnote{The exact beam energy for the Ru+Ru
collisions is 1.69$A$ GeV.} ($4\leq b \leq6$ fm), 2$A$ GeV Pb+Pb
($5\leq b \leq 8$ fm), 158$A$ GeV S+S ($3\leq b \leq 5$ fm) and 158$A$
GeV Pb+Pb ($5\leq b \leq 8$ fm) collisions. The impact parameter range
is chosen such that the maximum value of $v_1$ is achieved for any
given colliding system.

It is known from the hydrodynamic calculations that the pressure
created at the early stage will lead to sizable values of $v_1$ and
$v_2$, and the pressure is different for different bombarding
energies~\cite{Koch93}. However, from Figure~\ref{fig2} and
Table~\ref{table:v1v2}, one notices that the maximum values of $v_1$
for nucleons are independent of the beam energy for a given colliding
system. For a fixed beam energy, the maximum value of $v_1$ depends on
the size of the colliding nuclei. The larger the size, the higher the
maximum value of $v_1$. This may suggest that the initial geometry,
{\it i.e.} the size of the colliding nuclei and impact parameter,
determines the strength of $v_1$ near the beam (target) rapidity
region, providing the elastic scattering amplitude does not vary much
within energy range discussed here. Similar trends are also seen for
pions even though the $v_1$ for pions has the opposite sign to that of
nucleons due to the shadowing effect.

The other interesting feature is that the slope of the $v_1$ curve
around the mid-rapidity region depends on beam energy for both
nucleons and pions. The higher the beam energy, the smaller the
slope\footnote{This is also true for $v_1$ as a function of normalized
rapidity.}. The value of $v_2$, on the other hand, is maximized around
mid-rapidity. It was suggested that the second Fourier moment may
carry the information about the early stage of the
collision~\cite{Sorge97,Danielewicz98}. The sensitivity of the $v_2$
signal to the collision dynamics will be discussed in the next
section.


\section{ Time evolution of (\small{$v_1, v_2$})}

Now let us explore ($v_1, v_2$) of nucleons and pions as a function of
freeze-out time for two colliding systems: 2$A$ GeV Au+Au and 158$A$
GeV Pb+Pb collisions.  In a discussion of the time evolution, a brief
remark concerning the nuclear passing time is necessary. Nuclear
passing time is the time needed for two incoming nuclei to pass each
other and defined as $t_p=2r/(\gamma\beta)$ where $r$ is the nuclear
radius. Passing time sets the scale on which the effects of possible
nuclear shadowing, e.g. the interaction of the participating nucleons
and produced particles with the spectator matter, are
significant. Obviously, the passing time depends on the beam energy
and on the size of colliding nuclei. The nuclear passing time $t_p$,
in the nucleon-nucleon center-of-mass frame, for 2$A$ GeV Au+Au and
158$A$ GeV Pb+Pb collisions are listed in Table~\ref{table:pass}. Also
listed are the nucleon mean freeze-out time $t_m$ from the RQMD
calculations. The ratio of $t_m/t_p$ for the low energy collisions is
much smaller than that at the high energy, indicating that the nuclear
shadowing is relatively more important for lower energy collisions.

\subsection{\small{$v_1, v_2$} (momentum space)} 

\begin{center}
\begin{figure}[c]
\centerline{\psfig{file=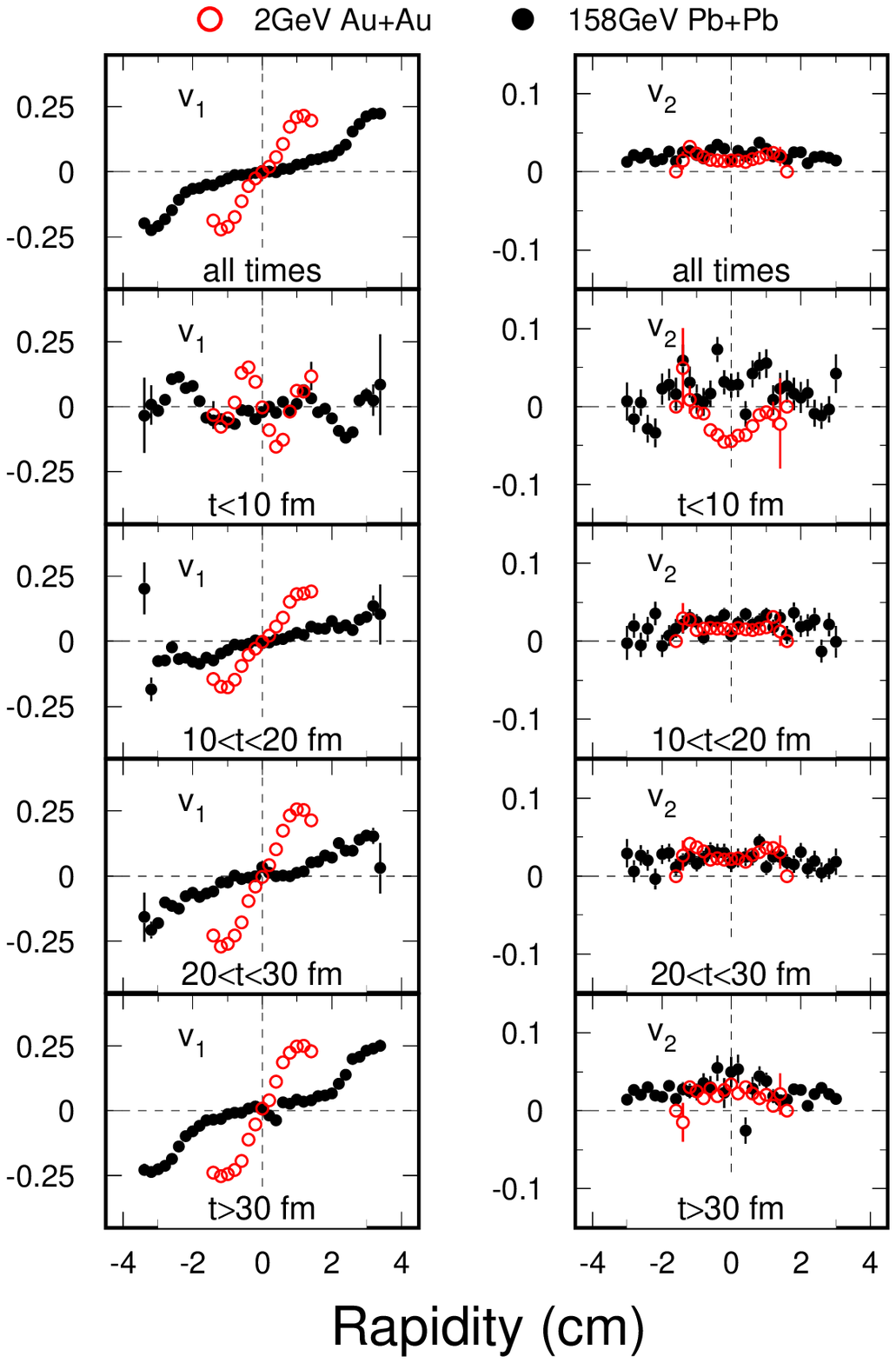,width=10.5cm}}
\caption{ The time evolution of $v_1$ and $v_2$ for nucleons in the
momentum space as a function of rapidity in the center-of-mass frame
for 2$A$ GeV Au+Au collisions (open circles) and 158$A$ GeV Pb+Pb
collisions (filled circles).}
\label{fig3}
\end{figure}
\end{center}

\begin{center}
\begin{figure}[c]
\centerline{\psfig{file=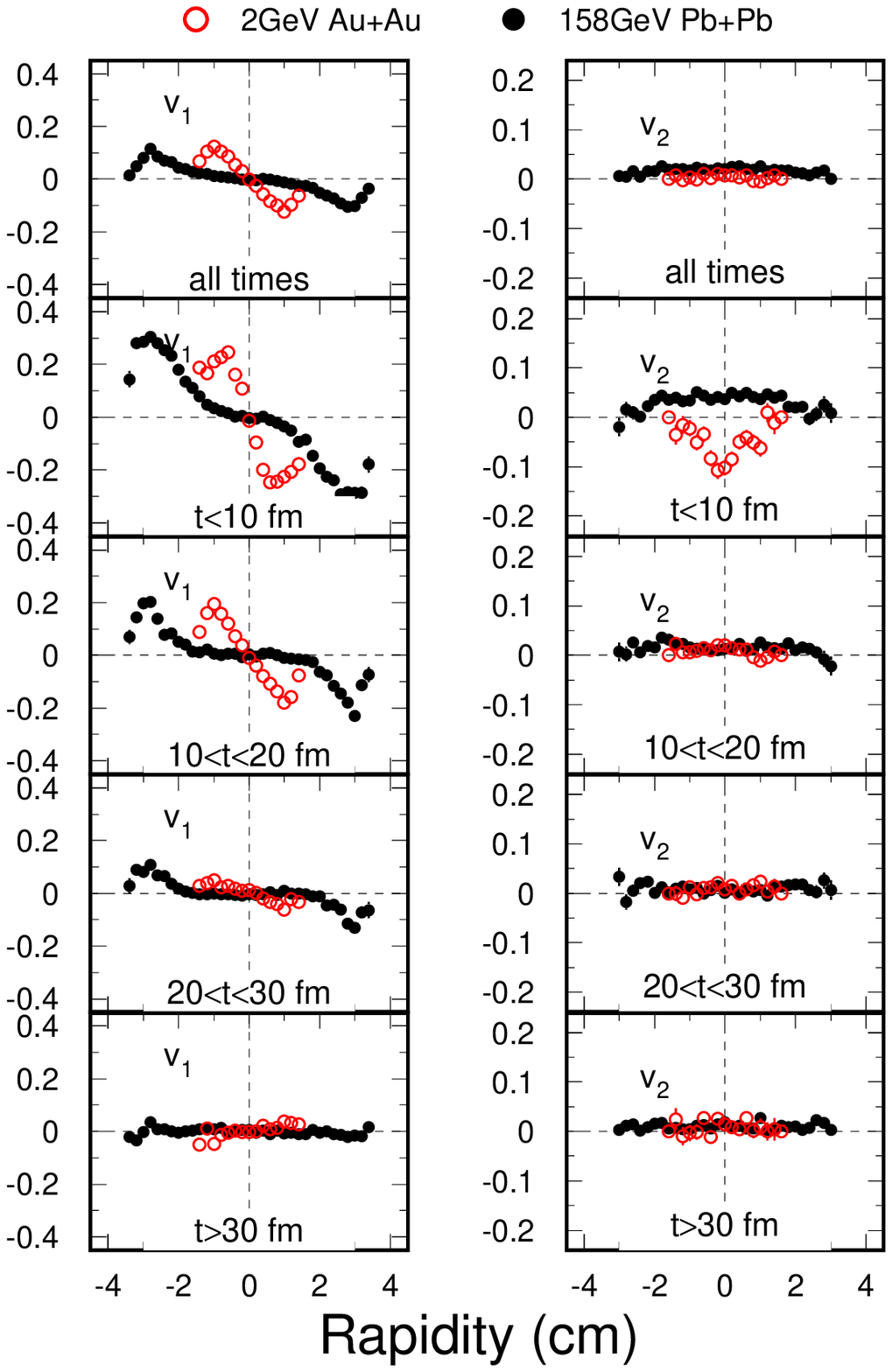,width=10.5cm}}
\caption{ The time evolution of $v_1$ and $v_2$ for pions in the
momentum space as a function of rapidity in the center-of-mass frame
for 2$A$ GeV Au+Au collisions (open circles) and 158$A$ GeV Pb+Pb
collisions (filled circles).}
\label{fig4}
\end{figure}
\end{center}

\begin{center}
\begin{figure}[c]
\centerline{\psfig{file=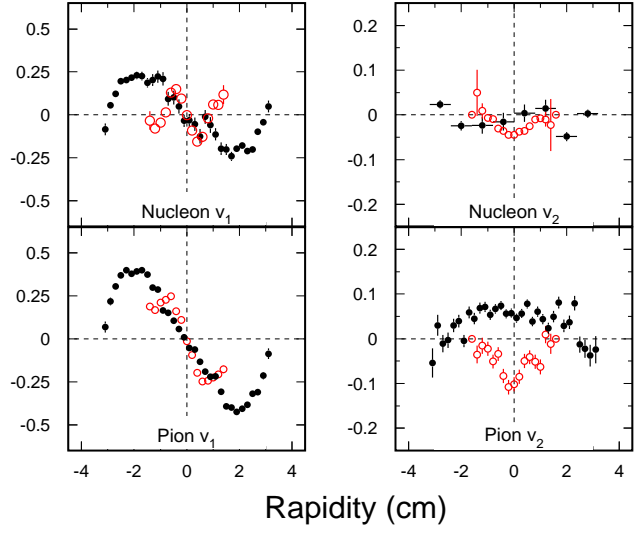,width=10.5cm}}
\caption{ Nucleon and pion anisotropic parameters $v_1, v_2$ as a
function of rapidity, around the nuclear passing time, for 2$A$ GeV
Au+Au (open circles) and 158$A$ GeV Pb+Pb (filled circles)
collisions. }
\label{fig5}
\end{figure}
\end{center}

The calculated ($v_1, v_2$) values, determined in the momentum space,
as a function of rapidity in different time windows are shown in
Figures~\ref{fig3} and \ref{fig4} for nucleons and pions,
respectively. In these figures, the open circles are for the lower
energy (2$A$ GeV) and the filled circles for the higher energy (158$A$
GeV) collisions. Similar distributions calculated at the corresponding
passing time ($0 \leq t \leq 12$ fm/c for 2$A$ GeV and $0 \leq t \leq
2$ fm/c for 158$A$ GeV collisions) are shown in Figure~\ref{fig5}.

For nucleons, as one can see in Figure~\ref{fig3} (left column lower 4
plots), the strength of $v_1$ is gradually built up as a function of
collision time. At the earlier time, $t \leq 10$ fm/c, a well defined
negative slope (around mid-rapidity) is seen for the low energy
collisions while at the high energy the distribution is relatively
flat. This flatness is due to the averaging over the 10 fm/c time
period which is much longer than the passing time for the 158$A$ GeV
collisions (see Table~\ref{table:pass}). In fact, the negative slope
is also present at that energy for $t \leq t_p$ as shown in
Figure~\ref{fig5}. This again suggests that the shadowing of the cold
nuclear matter is responsible for the observed negative slope at the
early stage of the collision. At $t > 10$ fm/c, the maximum value of
$v_1$ has been well established for the lower energy collisions. For
the higher energy collisions, the fully established $v_1$ only appears
after $t > 30$ fm/c.

The time dependence of $v_1$ for pions shows a different trend (Figure
4 left column). The maximum values of $v_1$ are achieved at the early
stage of the collision. As time goes on, the distributions of $v_1$
become flatter. The fact that the $v_1$ distributions for pions always
have a negative slope through the whole collision history suggests the
importance of shadowing effect. However, the shadowing matter is
different for pions at different collision stages. At early
time, $t \leq t_p$, pions (as well as the participating nucleons) are
shadowed by the cold spectators. Later, after the spectator matter
leaves the collision zone, pions are shadowed by the participant
nucleons. This may be the underlying mechanism that leads to the
different behavior of $v_1$ for nucleons and pions.

The second Fourier coefficient $v_2$ describes an elliptical shape of
the azimuthal distribution. A positive sign of $v_2$ means that the
longer axis of the ellipse is oriented in the reaction plane while a
negative sign of $v_2$ indicates the longer axis is in the direction
perpendicular to the reaction plane. This latter effect was called
``squeeze-out'' at Bevalac energies~\cite{Gutbrod90}. In
Figure~\ref{fig3} and Figure~\ref{fig4} (right column), the $v_2$
distributions are shown as a function of rapidity in different time
intervals for nucleons and pions, respectively. It can be seen that
the time averaged values of $v_2$ for nucleons and pions at both beam
energies are positive, implying that particles preferentially freeze-out
in the reaction plane.

However, at the time scale comparable with nuclear passing time,
nucleons and pions demonstrate an out-of-plane emission pattern
(``squeeze-out'') for 2$A$ GeV collisions (Figure~\ref{fig5} right
columns) which is consistent with the cold nuclear matter shadowing
scenario. For high energy collisions, the ``squeeze-out'' effect is
absent for both pions and nucleons. This may be explained by a much
shorter nuclear passing time and higher Lorentz contraction of the
colliding nuclei. According to the RQMD calculations, at about the
corresponding passing time, the average number of collisions suffered
by the mid-rapidity nucleons is $3$ and $5$ for 158$A$ GeV and 2$A$
GeV collisions, respectively. In the high energy collisions the
rapidity span is about 6 units. Hence, three collisions are just about
enough to move a nucleon from projectile rapidity to mid-rapidity. At
this energy, neither time nor number of rescatterings is sufficient to
develop an out-of-plane emission.

\subsection{\small{$v_1, v_2$} (configuration space)}

Up to now, we have discussed the anisotropy parameters ($v_1,v_2$) in
the momentum space. It is interesting to look at these parameters in
coordinate space in order to understand the space-time evolution of
the collisions within the model. To the best of our knowledge such
studies have not been performed before. The information about source
anisotropy may be obtained experimentally by studying two-particle
correlation functions~\cite{Voloshin97,Voloshin96,Urs98,Panitkin97}.

\begin{center}
\begin{figure}[c]
\centerline{\psfig{file=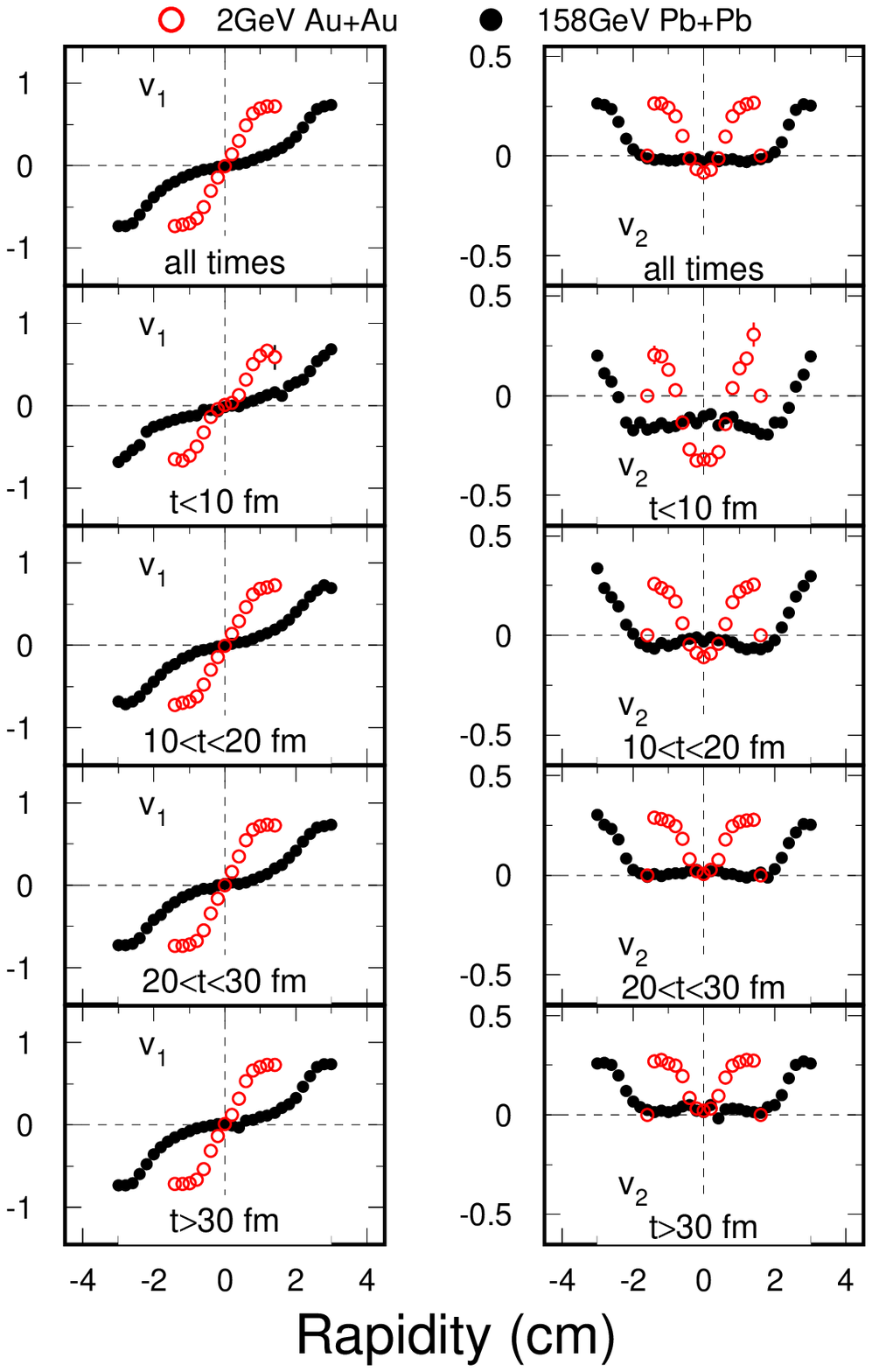,width=10.5cm}}
\caption{ The time evolution of $v_1$ and $v_2$ for nucleons in the
configuration space as a function of rapidity in the center-of-mass
frame for 2$A$ GeV Au+Au collisions (open circles) and 158$A$ GeV
Pb+Pb collisions (filled circles).}
\label{fig6}
\end{figure}
\end{center}

\begin{center}
\begin{figure}[c]
\centerline{\psfig{file=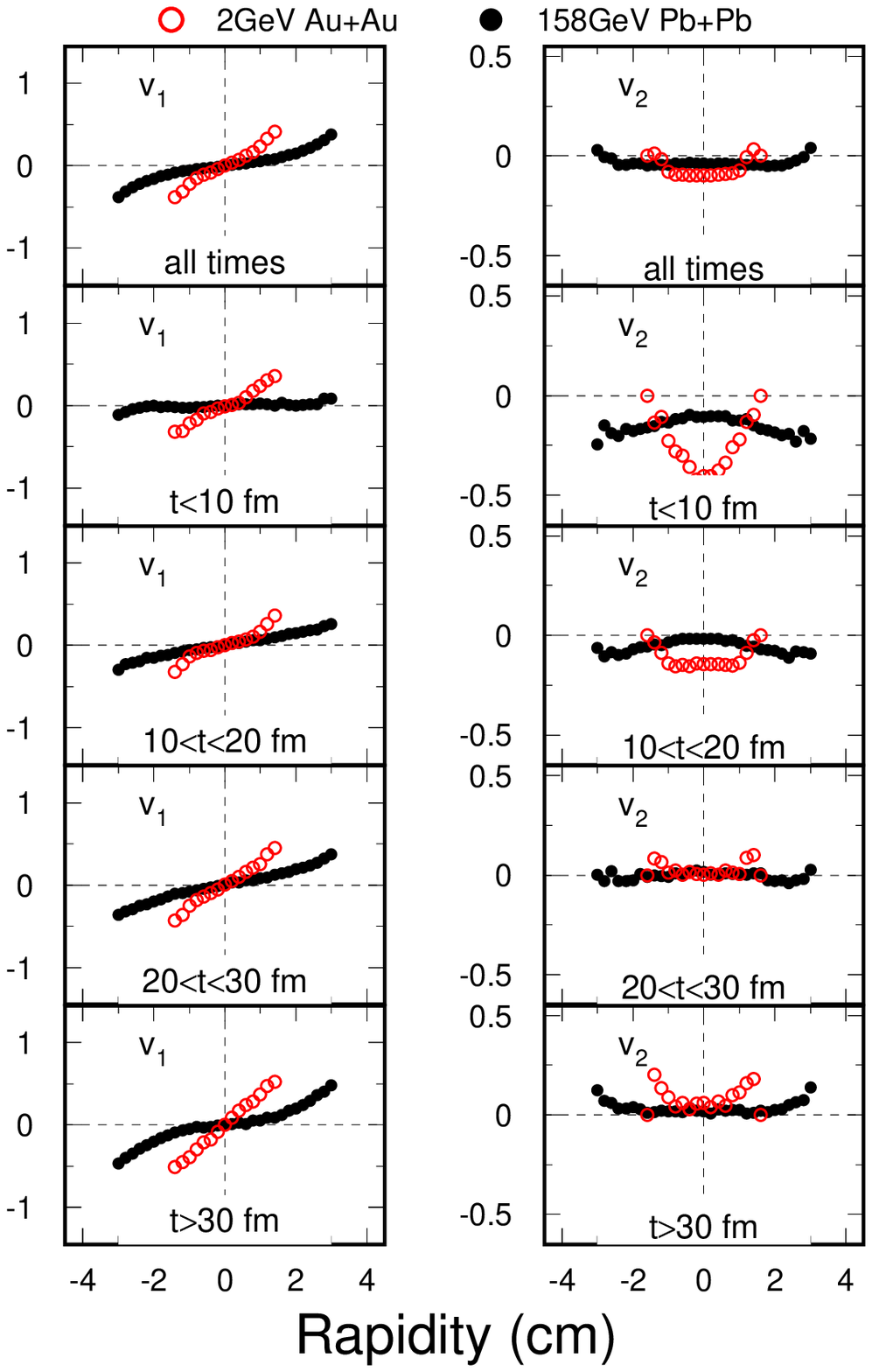,width=10.5cm}}
\caption{ The time evolution of $v_1$ and $v_2$ for pions in the
configuration space as a function of rapidity in the center-of-mass
frame for 2$A$ GeV Au+Au collisions (open circles) and 158$A$ GeV
Pb+Pb collisions (filled circles).}
\label{fig7}
\end{figure}
\end{center}

In general, the $v_1$ parameters in the configuration space for both
nucleons and pions (see Figure~\ref{fig6} and \ref{fig7} left columns)
behave differently from those in the momentum space: (1) The shapes of
the $v_1$ distributions are different and the maximum values of $v_1$
(around the beam and target rapidities) in the configuration space are
much larger than those in the momentum space; (2) The $v_1$ of pions
has the same sign to that of nucleons; (3) The $v_1$ becomes stronger
as a function of time, especially for pions.  Furthermore, similar to
what have been observed in the momentum space, the maximum values of
$v_1$ are the same for the two collisions at different beam
energies. This implies that the initial collision geometry plays an
important role here.
 
For the non-zero impact parameter heavy-ion collisions, the
transverse shape of the initial collision zone is an ellipse with its
longer axis oriented perpendicular to the reaction plane ($v_2 < 0$).
Such an event shape is indeed seen for both collision energies in
Figure~\ref{fig6} (right column) for the time interval $t\leq10$
fm/c. As time goes on, rescattering occurs among particles which leads
to a more spherical shape of the reaction zone, $v_2 \rightarrow
0$. Note one would expect a large positive $v_2$ at later times in
the case of pressure driven hydroexpansion. For the time integrated
distributions, some remnants of the initial shape still can be seen
for the 2$A$ GeV collisions while for the higher energy collision the
source appears to be isotropic.

\section{Summary}

Based on the RQMD(v2.3) model calculations, we studied the anisotropy
parameters ($v_1, v_2$) as a function of time for different energies
and different colliding systems. From a comparison between the model
calculations and the recent experiment measurement at the beam energy
of 158$A$ GeV, we find that the model provides a reasonable
description of the observed event anisotropy. From the study of energy
dependence, we find that the combination of collision geometry and
nuclear passing time plays a dominant role in determining the behavior
of the $v_1$ and $v_2$ parameters. Our studies show that the anisotropy
parameters extracted in the momentum space are different from the ones
extracted in the configuration space. As a function of collision time,
some rather dramatic changes are found for the event anisotropy
parameters in both momentum and configuration spaces. At $t \leq t_p$,
the pattern of the distributions is strongly influenced by the
spectators. At later times, a prominent in-plane emission is evident.

\vspace{0.2in}
 
{\bf Acknowledgments} We wish to express our gratitude to
Drs. J. Kapusta, D. Keane, V. Koch, A. Poskanzer, G. Rai, H.G. Ritter,
H. Sorge, H. St\"{o}cker, and S. Voloshin for many helpful
discussions.  We especially thank Dr. H. Sorge for providing the RQMD
code. This research used resources of the National Energy Research
Scientific Computing Center. This work has been supported by the
U.S. Department of Energy under Contract No. DE-AC03-76SF00098 and
National Science Foundation.
 


\begin{table}[htb]
\begin{tabular}{|c||c|c|c|c|}
Colliding System  &$A$ & r (fm) & y($v_1^{max}$) & $v_1^{max}$ \\ \hline
2$A$ GeV S+S   &32  & 3.8 & 0.91 & 0.171$\pm$0.007 \\ \hline
2$A$ GeV Ru+Ru &96  & 5.5 & 0.91 & 0.186$\pm$0.003 \\ \hline
2$A$ GeV Pb+Pb &208 & 7.1 & 0.91 & 0.225$\pm$0.009 \\ \hline\hline
158$A$ GeV S+S    &32  & 3.8 & 2.92 & 0.166$\pm$0.008 \\ \hline 
158$A$ GeV Pb+Pb  &208 & 7.1 & 2.92 & 0.225$\pm$0.009 \\
\end{tabular}
\vspace{0.2in}
\caption{\it Nucleon $v_1$ parameters for symmetric $A-A$
collisions, the atomic number $A$, the radius ($r=1.2\cdot A^{1/3}$) of
the beam(target) nucleus and the rapidity corresponding to the maximum
value of $v_1$ are shown.}
\label{table:v1v2}
\end{table}

\nopagebreak
\begin{table}[h]
\begin{tabular}{|c||c|c|c|c|c|}
Energy & $\beta$($c$) & $\gamma$ & $y$
& $t_p$ (fm/c) & $t_{m}$ (fm/c)\\ \hline
2~GeV & 0.718 & 1.438 &  0.904 & 13.52 & 20.63 \\ \hline
158~GeV & 0.994 & 9.231 & 2.913 & 1.55 & 33.89 \\
\end{tabular}
\vspace{0.2in}
\caption{ \it Velocity $\beta$ (in the unit of speed of light),
$\gamma$, rapidity $y$ and passing time $t_p$ are shown for 2$A$ GeV
Au+Au and 158$A$ GeV Pb+Pb collisions. The RQMD mean freeze-out time
$t_{m}$ for nucleons are also included. All quantities are evaluated
in the nucleon-nucleon center-of-mass frame. }
\label{table:pass}
\end{table}

\end{document}